\documentclass[aps,prl, 10pt, twocolumn]{article}%
\usepackage{amsfonts}
\usepackage[T2A]{fontenc}
\usepackage[cp1251]{inputenc}
\usepackage{amsmath}
\usepackage{amssymb}
\usepackage[russian,english]{babel}
\usepackage{graphicx}%
\begin{document}

\title{Thermal Hall Effect in 2D model for paramagnetic
dielectrics}
\author{L.A.Maksimov, T.V.Khabarova}
\date{\small{\textit{Kurchatov Institute, Moscow 123182, Russia}}}

\maketitle

\small

\begin{abstract}
Phonon polarization in a magnetic field is analyzed in 2D model.
It is shown, that at presence of spin-phonon interaction phonon
possess elliptic polarization which causes the appearance of heat
flux component perpendicular both to temperature gradient and
magnetic field.

\end{abstract}
\normalsize

\vspace{2mm}

\bigskip

Recently a magnetic field dependence of heat conductivity in
dielectric crystal $Tb_{3}Ga_{5}O_{12}$ has been experimentally
established \cite{strohm}, \cite{inushtald}. It is so called
Phonon Hall Effect (PHE) - a temperature gradient the has been
measured in a direction, perpendicular both to a heat flow and
magnetic field. This phenomenon is caused by spin-phonon
interaction (SPI) of phonons and paramagnetic ions. The theory of
this phenomenon was considered in works \cite{km}, \cite{sheng}.
We can expect similar effect to be detected in dielectrics where
molecules has rotary degrees of freedom. This effect is known for
a long time in gases as Senftleben-Beenakker effect
\cite{beenakker}. Generalization  of this phenomenon on molecular
crystals is considered in work  \cite{hb}. The theory of Hall
effect in ionic and molecular dielectrics for three-dimensional
case is rather cumbersome. Therefore it is interesting to consider
the theory in two-dimensional model. Probably, such theory can be
applied to some experiments on quasi-two-dimensional crystals
(films or surfaces of three-dimensional substances). The spectrum
and phonon polarization in presence of SPI was found. Expressions
for these polar vectors demonstrate nontrivial symmetry, which is
important for PHE theory construction. Transverse heat
conductivity coefficients and nondiagonal density matrix
calculations from \cite{km} for two-dimensional case are briefly
reproduced. In conclusion, we estimate the amount of effect and
discuss the results.

We consider dielectric consisting of light atoms and paramagnetic
particles (atoms or molecules) with magnetic moment $\textbf{M}$.
In the two-dimensional case we assume that $\textbf{M}$ is
perpendicular to 2D-crystal plane. For simplicity we will accept
that in a crystal cell there is one paramagnetic particle. The
magnetic moment of rare-earth atoms is caused by electrons of
$f$-subshell, the magnetic moment of molecules is proportional to
their rotary moment. In both cases the spectrum of paramagnetic
particles has a complex structure. But at low temperatures several
bottom levels play the main role, and in simplest approximation we
can replace the magnetic moment by pseudospin $\textbf{s}_{n}$
($n$ is a cell number). We will assume that  $T_{c}\ll T\ll
\Theta$, where  $T_{c}$ is a  rotary degrees of freedom freezing
temperature, $\Theta $\ -- Debye temperature. In these conditions
atomic oscillations in crystal are caused by long-wave acoustic
phonons. Thus all atoms in elementary cell oscillate with the same
amplitude $\textbf{U}_{n}$ and velocity  $\textbf{V}_{n}$, and the
general motion of atoms in a cell creates the total orbital moment
$[\textbf{U}_{n}\times \textbf{P}_{n}]$, где $\textbf{P}_{n}=m_{0}\textbf{V}%
_{n}$, and $m_{0}$ is the total mass of atoms in a cell.
Interaction between the moments of paramagnetic particles and the
orbital moment of a cell we will describe by the following
spin-phonon interaction Hamiltonian \cite{cap}:
\begin{equation*}
H_{1}=g\sum_{n}\left( \textbf{s}_{n},[\textbf{U}_{n}\times
\textbf{P}_{n}]\right) .
\end{equation*}%
In this work we use a system of units where  $k_{B}=1,\ \hbar =1$.
On the acoustic phonon wavelength scale there is a magnetization
self-averaging, and we can replace operator  $\textbf{s}_{n}$
with a value $\left\langle \textbf{s}%
\right\rangle =\left\langle \textbf{s}_{n}\right\rangle \sim \left\langle \textbf{M%
}_{n}\right\rangle $ averaged over a crystal. We consider the
interaction constant g as a phenomenological parameter. The value
of g strongly depends on crystal field and it was estimated in
numerous works \cite{cap}.

Let us find the acoustic phonon renormalization caused by this
interaction in a two-dimensional crystal. First of all, SPI
changes a relation between velocity and an momentum of oscillating
atoms (see \cite{km})
\begin{equation}
\textbf{V}_{n}=\frac{d\textbf{U}_{n}}{dt}=\frac{\partial H}{\partial \textbf{P}_{n}}=%
\textbf{P}_{n}/m_{0}+g[\left\langle \textbf{s}\right\rangle \times
\textbf{U}_{n}] \label{801}
\end{equation}
The oscillation equation in the momentum representation is
\begin{equation}
\omega _{ks}^{2}u_{ks}^{a}=\tilde{D}_{k}^{ab}u_{ks}^{b}
\label{157}
\end{equation}
Here $\textbf{u}_{ks}$ is the normalized polarization vector,
$\textbf{k}$ is the phonon wave vector, $s$ is a number of
acoustic mode, and
\begin{equation}
\tilde{D}_{k}^{ab}=D_{k}^{ab}-ie_{abc}G^{c},\ \   G^{c}=2\omega
g\left\langle s^{c}\right\rangle. \label{159}
\end{equation}
We see, that SPI adds an imaginary antisymmetric tensor to the
symmetric dynamic matrix $D_{k}^{ab}$. In the zero approximation
on SPI, the solution of (\ref{157}) determines the dispersion law
for two acoustic branches and two corresponding orthonormal
polarization vectors. The eigenvalue equation is
\begin{equation*}
\left( \omega _{ks}^{2}-D_{k}^{xx}\right) \left( \omega
_{ks}^{2}-D_{k}^{yy}\right) =\left( D_{k}^{xy}\right) ^{2}.
\end{equation*}%
To simplify, we assume crystal surface to be square-symmetric
\cite{ll7}
\begin{equation}
D_{k}^{ab}=A_{1}\delta ^{ab}k^{2}+A_{2}\delta
^{ab}k_{a}^{2}+A_{3}k_{a}k_{b}. \label{45}
\end{equation}%
It's easy to verify, that in isotropic model ($A_{2}=0$) the
longitudinal mode has the energy
$\omega_{k\parallel}=k\sqrt{A_{1}+A_{3}}$, and the transverse mode
energy is $\left( \omega_{0}=k\sqrt{A_{1}}\right) $. In the
general case we have
\begin{eqnarray}
\omega _{s}^{2} &=&\frac{1}{
2}(2A_{1}+A_{2}+A_{3})k^{2}+s\frac{1}{2}R,\ s=\pm
1,  \label{112} \\
R^{2} &=&(A_{2}+A_{3})^{2}(k_{x}^{2}-k_{y}^{2})^{2}+4\left(
A_{3}k_{x}k_{y}\right) ^{2}  \label{114}
\end{eqnarray}
For brevity, we omit the $\textbf{k}$ dependence indication in
evident cases here and below. In the general case, all of $A_{i}$
are positive and nonzero. So, in real 2D-crystal the acoustic
phonon spectrum is nondegenerated for all $\textbf{k}$ in zero
approximation, and $\omega _{k+}^{2}>\max
(D_{k}^{xx},D_{k}^{yy})$, $\omega _{k-}^{2}<\min
(D_{k}^{xx},D_{k}^{yy})$.

From (\ref{112}) and (\ref{114}) we see that $R^{2}$ and $\omega
_{\pm }^{2}$ are invariant with respect to inversion and following
reflections: $\left( k_{x},k_{y}\right) \rightarrow \left(
k_{x},-k_{y}\right) ,\ \left( k_{x},k_{y}\right) \rightarrow
\left( -k_{x},k_{y}\right)$. For heat flow calculation we also
need anisotropic velocity of sound expressions:
\begin{eqnarray*}
c_{ks}^{a}=\frac{\partial \omega _{s}}{\partial k^{a}}=\frac{1}{2\omega _{s}}\{(2A_{1}+A_{2}+A_{3})+\\+s\frac{1}{R}%
[(A_{2}+A_{3})^{2}(k_{a}^{2}-k_{a+1}^{2})
+2A_{3}^{2}k_{a+1}^{2}]\}k^{a}.
\end{eqnarray*}
If all of $A_{i}$ has the same order of magnitude, then both
velocities of sound has similar to $\textbf{k}$ direction, and $c
\simeq \sqrt{A}$ on the order of magnitude.

Polarization vectors play a great role in the PHE problem. From
dispersion equation (\ref{157}) we have
\begin{equation}
e_{ks}^{x}=\left( \omega _{ks}^{2}-D_{k}^{yy}\right)
C_{ks}signk^{x},\ e_{ks}^{y}=\left\vert D_{k}^{xy}\right\vert
C_{ks}signk^{y},  \label{1}
\end{equation}
and normalization
\begin{equation*}
C_{ks}^{-2}=\left( \omega _{ks}^{2}-D_{k}^{yy}\right)
^{2}+D_{k}^{xy2}=sR\left( \omega _{ks}^{2}-D_{k}^{yy}\right) >0
\end{equation*}%
Eigenvectors (\ref{1}) and $C_{ks}$ are determined to phase, which
depends on $\textbf{k}$ and $s$. Let us take $C_{ks}=\left\vert
C_{ks}\right\vert $. Then normalized polarization vectors are real
and can be written down as follows:
\begin{equation}
\begin{array}
[c]{c}%
e_{ks}^{x}=s\ast signk^{x}\sqrt{\frac{s\left( \omega
_{ks}^{2}-D_{k}^{yy}\right) }{R}},\ \\
e_{ks}^{y}=signk^{y}\sqrt{\frac{s\left( \omega
_{ks}^{2}-D_{k}^{xx}\right) }{R}}.
\end{array}
\label{5}
\end{equation}
They also possess properties:
\begin{equation}
\begin{array}
[c]{c}%
\left( \textbf{e}_{+}\textbf{e}_{-}\right) =0,\ \ \left( \textbf{e}_{+}\times \textbf{e}%
_{-}\right) ^{z}=C_{k+}C_{k-}D_{k}^{xy}R,\ \left(
\textbf{e}_{s}\hat{k}\right) =\\
=s\ast \sqrt{\frac{\left\vert k^{x}\right\vert
^{2}}{k^{2}}\frac{s\left( \omega _{ks}^{2}-D_{k}^{yy}\right)
}{R}}+\sqrt{\frac{\left\vert k^{y}\right\vert
^{2}}{k^{2}}\frac{s\left( \omega _{ks}^{2}-D_{k}^{xx}\right)
}{R}}.
\end{array}
\label{90}
\end{equation}
From the last equation in (\ref{90}) we can see that the upper
(lower) branch is approximately longitudinal (transverse).
Components (\ref{5}) changes their signs after inversion and
corresponding components changes their signs after reflection from
general axes as $\textbf{e}_{ks}$ are polar vectors. If we will
turn $\textbf{k}$ from $\varphi = 0$ to 2$ \pi$ smoothly, the
vector $\textbf{e}_{k-}$ changes its direction by a jump when k
crosses one of the axes. Vector $\textbf{e}_{k+}$ oscillates near
wave vector $\textbf{k}$ direction, vectors $\textbf{e}_{k+}$ and
$\textbf{e}_{k-}$ are always perpendicular to each other.

Let us now discuss phonon renormalization in linear approximation
on SPI. We present the polarization vector in linear approximation
in the form of $u_{s}^{a}=\left( e_{s}^{a}+\delta e_{s}^{a}\right)
$, and rewrite dispersion equation (\ref{157}) using this form.
Then from the real part of obtained equation we can see that
phonon spectrum and group velocity $c_{ks}=\partial \omega
_{ks}/\partial k$ are not renormalized ($\delta \omega _{s}=0$).
Imaginary part of dispersion equation determines the
renormalization of polarization vector:
\begin{equation*}
\left( \omega _{s}^{2}\delta ^{ab}-D^{ab}\right) \delta
e_{s}^{b}=-ie_{abc}G^{c}e_{s}^{b}.
\end{equation*}%
The most general form of the solution would have a form $\delta
e_{s}^{b}=iKe_{s^{\prime }}^{b}$, then we obtain
\begin{equation}
K=\left( \omega _{k+}^{2}-\omega _{k-}^{2}\right) ^{-1}\left( \left[ \textbf{e}%
_{k+}\times \textbf{e}_{k-}\right] \textbf{G}\right) .
\label{113}
\end{equation}

Thus, SPI leads to elliptic polarization of phonons, which is
expressed in the imaginary addition to polarization vectors in
zero approximation (\ref{5}). Using relation (\ref{159}) between
$\textbf{G}$ and $g$, for thermal phonons we can estimate the
degree of elliptic polarization:
\begin{equation}
K\simeq g\left\langle s\right\rangle /T.  \label{117}
\end{equation}%

As noted above, phonon spectrum and group velocity do not depend
on SPI in linear approximation. But the form of heat flow density
$j^{\gamma }$ in crystal can change in presence of SPI. However,
it can be shown that, due to linear relation between velocity and
momentum (\ref{801}), the expression for heat flow in coordinate
representation has exactly the same form as in \cite{hardy}
\begin{equation}
j^{c}=\frac{1}{2V}m_{0}\sum_{i\neq j}R_{ij}^{c}D_{ij}^{ab}\left(
U_{i}^{a}V_{j}^{b}\right) ,  \label{2}
\end{equation}
but here $\textbf{V}_{n}=\textbf{P}_{n}/m_{0}+g[\textbf{s}\times
\textbf{U}_{n}]$ (see. (\ref{801})). Then we write (\ref{2}) in
secondary quantization representation, average it over the system
state, omit the anomalous averages $\left\langle
a_{ks}a_{-ks^{\prime }}\right\rangle $, $\left\langle
a_{-ks^{\prime }}^{+}a_{ks}^{+}\right\rangle $, change symbols
under the summation sign and obtain the following expression for
the heat flow density:
\begin{equation}
\begin{array}
[c]{c}%
\left\langle j^{c}\right\rangle =\frac{1}{4V}\operatorname{Re}\{\sum_{kss^{\prime }}(%
\sqrt{\frac{\omega _{ks}}{\omega _{ks^{\prime
}}}}+\sqrt{\frac{\omega _{ks^{\prime }}}{\omega _{ks}}})\left(
\nabla _{k}^{c}D_{k}^{ab}\right) \times
\\
\times u_{ks}^{a\ast }u_{ks^{\prime }}^{b}\rho _{ss^{\prime
}}(k)\},
\end{array}
\label{3}
\end{equation}
where $\rho _{ss^{\prime }}(k)=\left\langle
a_{ks}^{+}a_{ks^{\prime }}\right\rangle$ is phonon density matrix.
In zero approximation (\ref{3}) has the usual form of energy flow
density for phonon gas with $ n_{ks}=\left\langle
a_{ks}^{+}a_{ks}\right\rangle $. Substituting polarization vector
in linear approximation, we obtain (see \cite{km})
\begin{equation}
\begin{array}
[c]{c}%
\left\langle \delta j^{c}\right\rangle =\frac{2}{V}\sum_{k}(\sqrt{\frac{%
\omega _{k-}}{\omega _{k+}}}+\sqrt{\frac{\omega _{k+}}{\omega _{k-}}}%
)K(k)\omega _{k-}\omega _{k+} \times
\\
\times (c _{k+}^{c }-c _{k-}^{c })\operatorname{Im}\rho _{-+}(k).
\end{array}
\label{2010}
\end{equation}
If magnetic field and moment $\textbf{M}$ are directed along $z$
axis, temperature gradient $\nabla T$ is directed along $x$ axis,
and transverse heat flow is directed along $y$ axis, then,
substituting $\rho _{-+}$ in the form of linear response $\rho
_{-+}(k)=-iA_{-+}^{x}\left( k\right) \left( \nabla T\right) _{x}$
into (\ref{2010}), we have
\begin{equation}
\begin{array}
[c]{c}%
\varkappa ^{yx}=\frac{2}{V}\sum_{k}(\sqrt{\frac{\omega _{k-}}{\omega _{k+}}}+%
\sqrt{\frac{\omega _{k+}}{\omega _{k-}}})K(k)\omega _{k-}\omega
_{k+}\times \\
\times(c _{k+}^{y}-c _{k-}^{y})\operatorname{Re}A_{-+}^{x}(k),
\end{array}
\label{10}
\end{equation}
where $K(k)$ is determined in (\ref{113}). As $K(k)\sim
\textbf{G}$, then for calculation (\ref{10}) in linear
approximation, it would suffice to calculate $A_{12}^{x}(k)$ in
zero approximation. According to \cite{km}
\begin{eqnarray}
A_{pq}^{x}=\frac{F_{ss^{\prime
}}(\textbf{k})c_{s}^{x}(\textbf{k})+F_{s^{\prime
}s}(\textbf{k})c_{s^{\prime }}^{x}(\textbf{k})}{\omega _{ks}-\omega _{ks^{\prime }}%
}, \label{3222}
\\ F_{ss^{\prime }}(\textbf{k})=-\frac{\omega _{ks}\Omega
_{pq}}{2T^{2}\Omega _{pp}}N_{ks}\left( 1+N_{ks}\right) .
\label{322}
\end{eqnarray}%
Here $\Omega _{pq}$ are the effective relaxation frequencies,
$p=\textbf{k}s,\ q=\textbf{k}s^{\prime }$. In general case $\Omega
_{pq},\ \Omega _{qp},\ \Omega _{pp}$ have the same order of
magnitude (but $\left\vert \Omega _{pq}\right\vert ,\left\vert
\Omega _{qp}\right\vert <\Omega _{pp}$). In the two-dimensional
case the longitudinal component of heat conductivity tensor is
$\varkappa ^{xx}\simeq T^{2}\Omega ^{-1}$, where $c, \Omega $ are
the average values of $c_{p},\Omega _{pp}$. The substitution of
eq. (\ref{3222}) and (\ref{322}) into (\ref{10}) gives us the
transverse component in the form of integral dependent on the
relaxation frequencies ratio $\ \Omega _{pq}/\Omega _{pp}$.

The product $c^{x}c^{y}$ enter into the integral (\ref{10}),
instead of $\left( c^{x}\right) ^{2}$ in longitudinal heat
conductivity, and it seems to disappear when averaging over
$\textbf{k}$ directions. However, in (\ref{10}) there is $K(k)
\sim \left( \textbf{e}_{1}\times \textbf{e}_{2}\right) \sim
D^{xy}\sim k_{x}k_{y}$. Therefore, under summation sign in
(\ref{10}) there is an expression invariant by inversion in
reciprocal space, and averaging over $\textbf{k}$ gives a nonzero
result. We note, that the presence of $(\left(
\textbf{e}_{1}\times \textbf{e}_{2}\right) \textbf{G})$ points to
determining role of the phonon elliptic polarization, which arises
due to SPI. The integrand in (\ref{10}) becomes maximum when
acoustic modes becomes at most close to each other. In 2D crystal
(when all $A_{i}$ in (\ref{45}) are the same order of magnitude)
there are no such preferential directions.

In contrast to 3D case \cite{km} the PHE magnitude (\ref{10}) is
strongly definable
\begin{eqnarray*}
\varkappa ^{yx} &=&\frac{2}{V}\sum_{k}(\sqrt{\frac{\omega _{k-}}{\omega _{k+}%
}}+\sqrt{\frac{\omega _{k+}}{\omega _{k-}}})K(k)\omega _{k-}\omega
_{k+}\times \\
&&\times(C _{k+}^{y}-C _{k-}^{y}) \frac{F_{ss^{\prime }}(\textbf{k})c_{s}^{x}(\textbf{k})+F_{s^{\prime }s}(\textbf{k}%
)c_{s^{\prime }}^{x}(\textbf{k})}{\omega _{ks}-\omega _{ks^{\prime
}}},
\end{eqnarray*}%
and can be calculated exactly. However, the problem definition has
a model character and we will make the estimations only. We assume
the collision frequencies $\Omega _{pq}$ and $\Omega _{pp}$ to be
the same order of magnitude $\Omega $. For simplicity we also
accept a certain average value for velocity of sound $\bar{c}$ for
both modes, and thermal phonons with frequencies about $T$ play a
main role in integral. In comparison to the longitudinal heat
conductivity, the transverse heat conductivity has two additional
multipliers: the degree of ellipticity (\ref{117}) and $\Omega
_{pq}/\left( \omega _{ks}-\omega _{ks^{\prime }}\right) $. They
characterize the magnitude of nondiagonal elements of density
matrix. The Hall angle has the following order of magnitude в
\begin{equation*}
\varkappa ^{yx}/\varkappa ^{xx}\simeq \left( \frac{g\left\langle
s\right\rangle }{T}\right) \left( \frac{\Omega }{T}\right).
\end{equation*}
The last multiplier can be estimated from experimental value of
heat conductivity coefficient. The first multiplier depends on SPI
magnitude and was discussed in works \cite{km}, \cite{cap},
\cite{splatt} for ionic dielectrics. We suppose this ratio to have
a comparable order of value in dielectrics containing molecules
with rotational degrees of freedom. It can be even greater due to
a higher degree of molecules anisotropy. Anyway, the transition
from 3D to 2D case can make easier the observation of PHE.

\end{document}